\documentclass[a4paper]{jpconf}
\usepackage{graphicx}
\usepackage[]{sidecap}
\usepackage{subfig}
\usepackage{cite}                               
\usepackage{caption}
\usepackage{textcomp}

\makeatletter
\let\ns\@undefined                        
\let\ms\@undefined                  
\makeatother

\usepackage[pdftex,
	    plainpages=false,
	    pagebackref=true]{hyperref}      
\usepackage[all]{hypcap}                     
\usepackage{amsmath}                         
\usepackage{amssymb}                         
\usepackage{xkeyval}                         %
\usepackage{xspace}                          %
\usepackage[alsoload=hep,
	    alsoload=prefixed,
	    alsoload=abbr]{siunitx}          

\hypersetup{
     unicode=false,          
     pdftoolbar=true,        
     pdfmenubar=true,        
     pdffitwindow=true,       
     pdftitle={Ground-state proton decay of 69Br and implications for the rp-process 68Se waiting-point},    
     pdfauthor={A M Rogers},     
     pdfsubject={},          
     pdfnewwindow=true,      
     pdfkeywords={rp process, 69Br, NSCL, X-ray Bursts, radioactivity}, 
     colorlinks=true,        
     linkcolor=black,        
     citecolor=blue,         
     filecolor=magenta,       
     urlcolor=blue           
}

\DeclareGraphicsExtensions{.pdf,.png,.jpg}
\graphicspath{{Figures/}}

\newcommand{\ie}{\textit{i.e.}}
\newcommand{\eg}{\textit{e.g.}}

\newcommand{\rp}{\textit{rp}~process}

\begin{document}
\title{Ground-state proton decay of $^{69}$Br and implications for the \textit{rp}-process $^{68}$Se waiting-point}

\author{A~M~Rogers$^{1,2,4}$, 
	W~G~Lynch$^{2,3,4}$,
	M~A Famiano$^{2,4,5}$,
	M~S~Wallace$^{2,6}$,
	F~Amorini$^{7}$,
	D~Bazin$^{2}$,
	R~J~Charity$^{8}$,
	F~Delaunay$^{2,9}$,
	R~T~de~Souza$^{10}$,
	J~Elson$^{8}$,
	A~Gade$^{2}$,
	D~Galaviz$^{2,4}$,
	S~Hudan$^{10}$,
	J~Lee$^{2}$,
	S~Lobostov$^{11}$,
	S~Lukyanov$^{11}$,
	M~Mato\v{s}$^{2,4}$,
	M~Mocko$^{2}$,
	M~B~Tsang$^{2,3}$,
	D~Shapira$^{12}$,
	~L~G~Sobotka$^{8}$,
	and~G~Verde$^{7}$
}

\address{$^{1}$ Physics Division, Argonne National Laboratory, Argonne, IL, 60439, USA}
\address{$^{2}$ National Superconducting Cyclotron Laboratory, Michigan State University, East Lansing, MI 48824, USA}
\address{$^{3}$ Dept. of Physics \& Astronomy, Michigan State University, East Lansing, MI 48824, USA}
\address{$^{4}$ JINA, Michigan State University, East Lansing, MI 48824, USA}
\address{$^{5}$ Dept. of Physics, Western Michigan University, Kalamazoo, MI 49008,USA}
\address{$^{6}$ Los Alamos National Laboratory, Los Alamos, NM 87545, USA}
\address{$^{7}$ Istituto Nazionale di Fisica Nucleare, Sezione di Catania, I-95123, Italy}
\address{$^{8}$ Dept. of Chemistry, Washington University, St. Louis, MO 63130, USA}
\address{$^{9}$ LPC Caen, ENSICAEN, Universit\'{e} de Caen, CNRS/IN2P3, Caen, France}
\address{$^{10}$ Indiana University Cyclotron Facility and Dept. of Chemistry, Bloomington, IN 47405, USA}
\address{$^{11}$ FLNR/JINR, 141980 Dubna, Moscow region, Russian Federation}
\address{$^{12}$ Oak Ridge National Laboratory, Oak Ridge, TN 37831, USA}

\ead{amrogers@phy.anl.gov}

\begin{abstract}
The first direct measurement of the proton separation energy, $S_{p}$, for the proton-unbound nucleus $^{69}$Br is reported.
Of interest is the exponential dependence of the 2$p$-capture rate on $S_p$ which can bypass the ${}^{68}$Se waiting-point in the astrophysical \rp.
An analysis of the observed proton decay spectrum is given in terms of the ${}^{69}$Se mirror nucleus and the influence of $S_p$ is explored within the context of a single-zone X-ray burst model.
\end{abstract}

\section{Introduction}
\begin{SCfigure}
 \centering
 \includegraphics[width=0.6\textwidth]{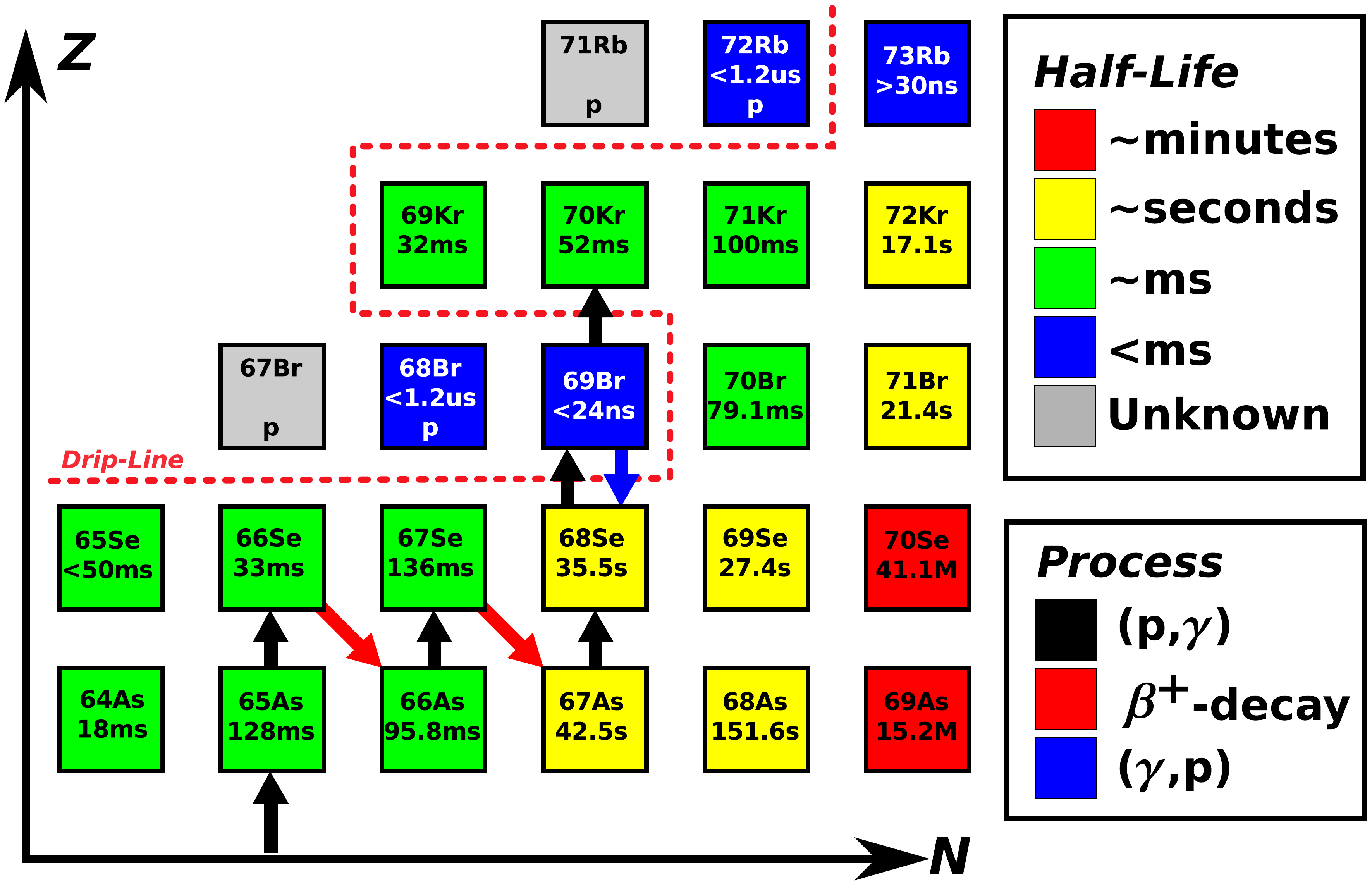}
 \caption{Illustration of the \textit{rp}-process ${}^{68}$Se waiting-point.  Proton-capture reactions, which are significantly faster than the respective $\beta$ decays, process nuclei until the proton drip-line is reached.  When the process reaches ${}^{68}$Se proton capture remains inhibited due to the unbound nature of ${}^{69}$Br.  Further processing must then wait for the long ${}^{68}$Se $\beta$ decay, essentially terminating the process.  However, 2$p$-capture reactions could bypass the waiting point.}
 \label{fig:chart}
\end{SCfigure}

The astrophysical rapid proton capture, or \rp, is characterized by proton-capture reactions along the proton drip line which occur faster than their subsequent $\beta$ decays.
The hydrogen-rich environment of type-I X-ray bursts is expected to be one of the key astrophysical sites where the \rp\ occurs, driving energy generation and nucleosynthesis during the cooling phase of the burst~\cite{taam1993successive}.
For nuclei where proton capture is inhibited (\eg\ at the proton drip line) and the $\beta$-decay half-life is long compared to the X-ray burst timescale of $\sim10-\SI{100}{s}$, a so-called \textquotedblleft waiting point\textquotedblright\ is reached.
One of the major waiting points occurs at ${}^{68}$Se ($T_{1/2}=\SI{35.5}{s})$ as illustrated in \autoref{fig:chart}.
It has been suggested, however, that the sequential two-proton capture reaction ${}^{69}$Br(2$p$,$\gamma$)${}^{70}$Kr can bypass the waiting point, allowing the process to continue onward to heavier masses~\cite{schatz1998rpn}.

The 2$p$-capture rate depends exponentially on the ${}^{69}$Br proton separation energy which is poorly constrained and has been determined only through indirect and theoretical methods~\cite{PhysRevLett.66.1571,Audi1995409,PhysRevLett.74.4611,PhysRevC.53.1753,PhysRevC.55.2407,PhysRevC.65.044618,PhysRevC.65.045802,Wohr2004349,Audi2003337,clark2004,schury:055801,savory:132501}.
The most recent non-observation measurement~\cite{PhysRevC.53.1753} has established a lifetime limit of $<\SI{24}{ns}$ and, consequently, a prediction of $S_{p}<\SI{-500}{keV}$.
Recent mass measurements of ${}^{68}$Se~\cite{savory:132501} and ${}^{69}$Se~\cite{schury:055801} using a precision penning trap combined with a Coulomb displacement energy (CDE) calculation~\cite{PhysRevC.65.045802} have yielded a value of $S_{p}=\SI{-606(105)}{keV}$ where the uncertainty is dominated by the theoretical CDE estimate.
To resolve the uncertainties in the proton separation energy we have performed an experiment to measure proton-unbound states through a complete kinematic reconstruction of the ${}^{69}$Br$\rightarrow p+{}^{68}$Se decay.

\section{Experimental details}
\begin{figure}
 \centering
  \includegraphics[width=0.95\textwidth]{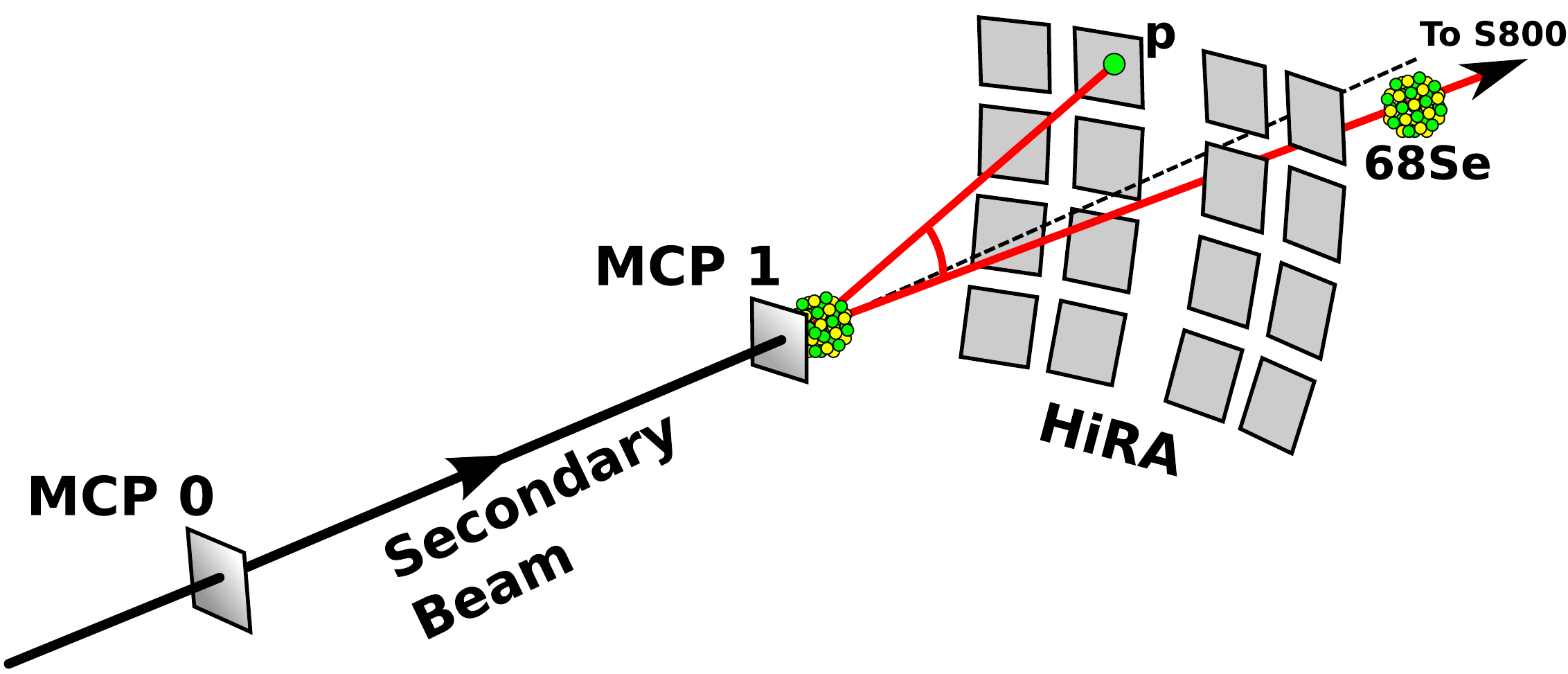}
  \caption{Schematic drawing of the experimental setup.  The incoming secondary beam, tracked by two MCPs separated by \SI{505}{mm}, interacts with the reaction target producing ${}^{69}$Br, which decays at the target location.  The decay proton momenta and scattering angle are measured in one of 16 HiRA telescopes while the heavy beam-like residue is measured at the S800 focal plane.}
  \label{fig:reaction}
\end{figure}

The experiment was conducted using the Coupled Cyclotron Facility at the National Superconducting Cyclotron Laboratory at Michigan State University (MSU).
A primary beam of \SI{140}{MeV/u} ${}^{78}$Kr was fragmented on a ${}^{9}$Be production target to produce a secondary cocktail beam, selected using the A1900 fragment separator~\cite{morrissey1997nhr} and composed primarily of $^{69}$As (\SI{23.9}{\percent}), $^{70}$Se (\SI{66.7}{\percent}), and $^{71}$Br (\SI{9.4}{\percent}).
The beam was transported to the experimental setup, shown schematically in \autoref{fig:reaction}, at the entrance of the S800 magnetic spectrograph~\cite{bazin2003ss,yurkon1999fpd} where it impinged on a \SI{5.4}{mg/cm^2} polypropylene $(\textrm{C}_{3}\textrm{H}_{6})_{n}$ reaction target.

Three detector systems were used to identify the ${}^{69}$Br$\rightarrow p+{}^{68}$Se coincident, forward-focused decay products, measure their final momenta, and reconstruct the relative energy of the two-body decay in the center-of-mass frame.
Protons were measured using 16 telescopes from the MSU high-resolution array (HiRA)~\cite{wallace2007hra}.
Each telescope consists of a \SI{1.5}{mm} thick $32\times32$ strip double-sided silicon detector providing a $\Delta E$ signal, which is backed by four \SI{4}{cm} $E$ CsI(Tl) detectors.
To correct for the uncertainty in the measured proton angle in HiRA \textthreequartersemdash\ caused by the large beam emittance \textthreequartersemdash\ two micro-channel plate detectors (MCPs), separated by \SI{505}{mm}, were used to track the incoming beam onto the target~\cite{Shapira2000396,Shapira2000409}.
The heavy beam-like ${}^{68}$Se was identified, and momentum analyzed using the S800 spectrograph.

%

\begin{figure}
 \centering
 \subfloat[]{\label{fig:s800Pid}\includegraphics[width=0.51\textwidth]{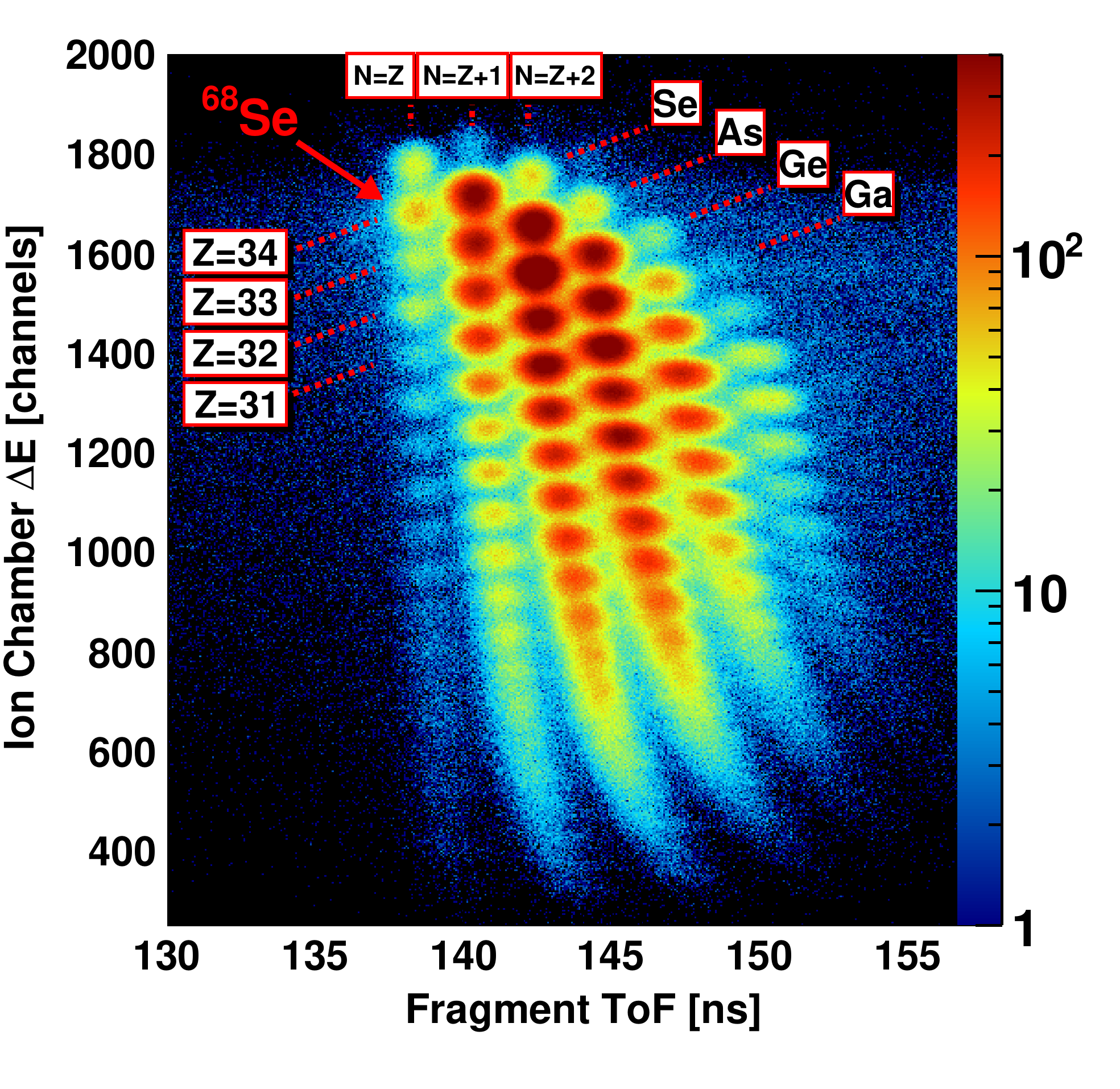}}
 \subfloat[]{\label{fig:hiraPid}\includegraphics[width=0.49\textwidth]{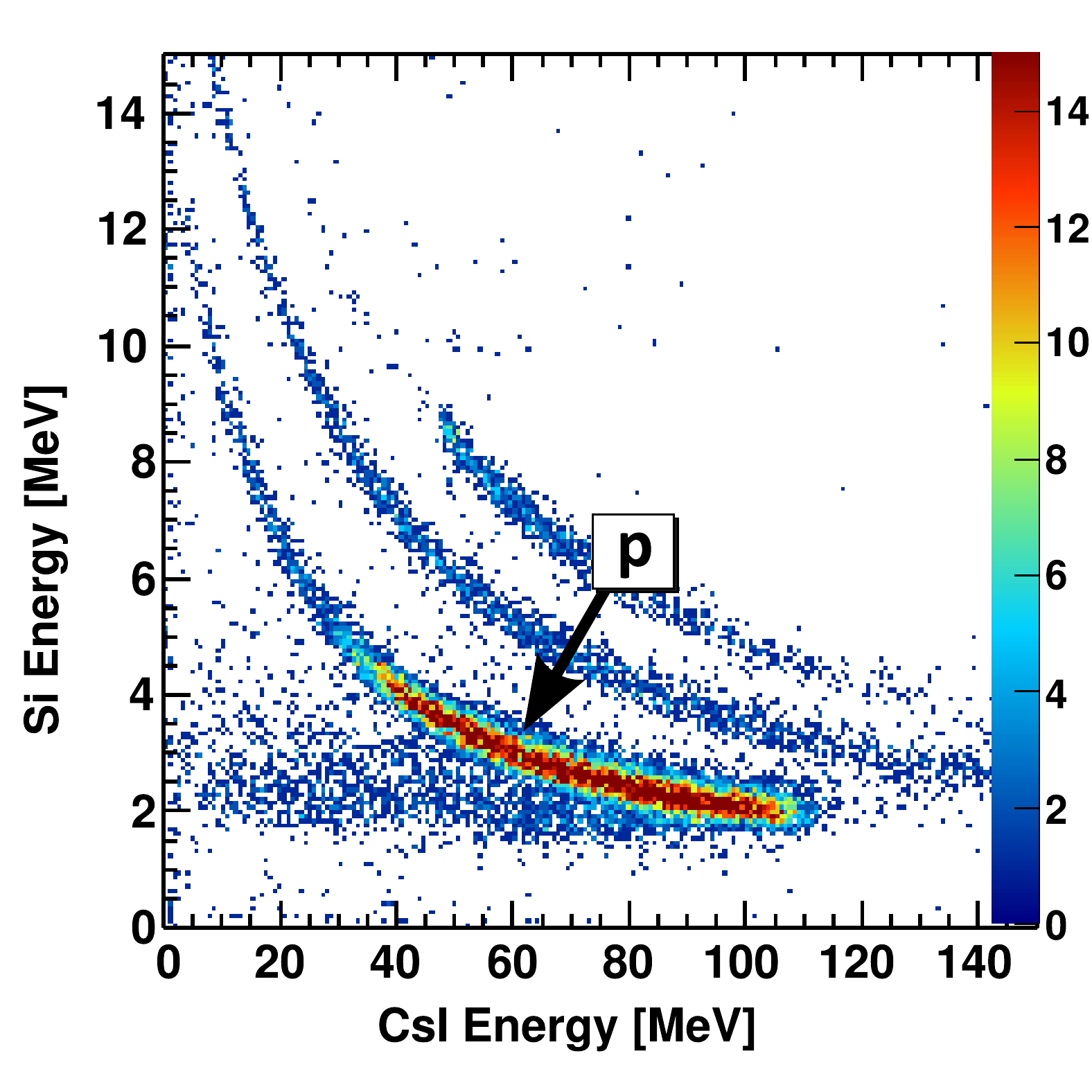}}
 \caption{(a) Particle identification spectrum of the beam-like residues detected at the S800 focal plane, in coincidence with single protons in HiRA, and produced in reactions with the $^{70}$Se secondary beam. (b) Particle identification spectrum of protons detected in HiRA in coincidence with ${}^{68}$Se at the S800 focal plane.}
 \label{fig:pid}
\end{figure}

Particle identification spectra from the S800 and HiRA, and produced in reactions with the ${}^{70}$Se beam, are shown in Fig.\autoref{fig:s800Pid} and Fig.\autoref{fig:hiraPid}, respectively.
Identification in the S800 is accomplished using the $\Delta E$-ToF method where the $\Delta E$ signal is taken from a segmented ionization chamber at the focal plane while the ToF is the time from the target to the focal plane.
A similar method using ToF-ToF was used to identify the incoming secondary beam.
An intense proton band coincident with ${}^{68}$Se events is observed in Fig.\autoref{fig:hiraPid}.

\section{Results and discussion}
\begin{figure}
 \centering
 \subfloat[]{\label{fig:ERelSpec}\includegraphics[width=0.55\textwidth]{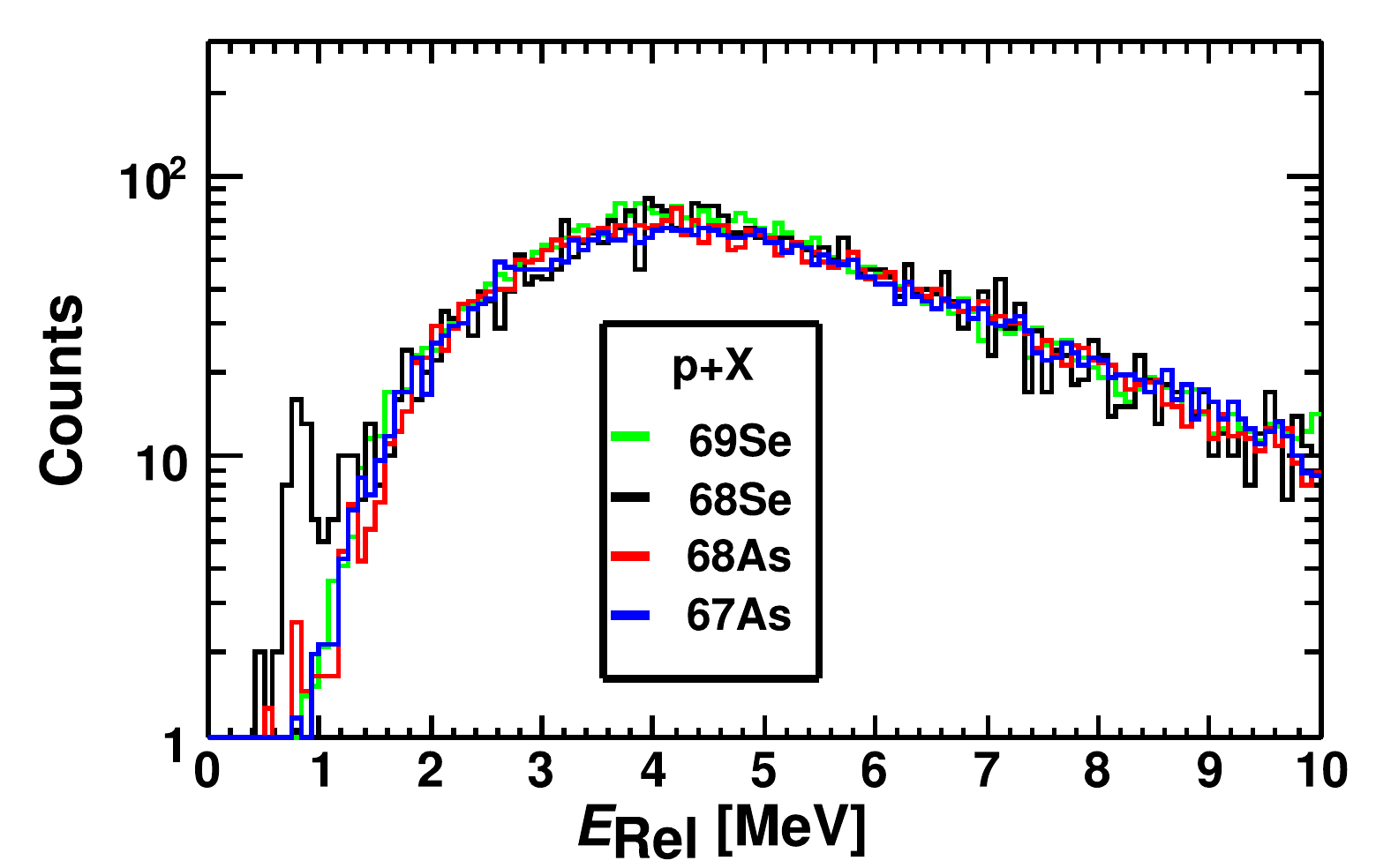}}
 \subfloat[]{\label{fig:Fit}\includegraphics[width=0.45\textwidth]{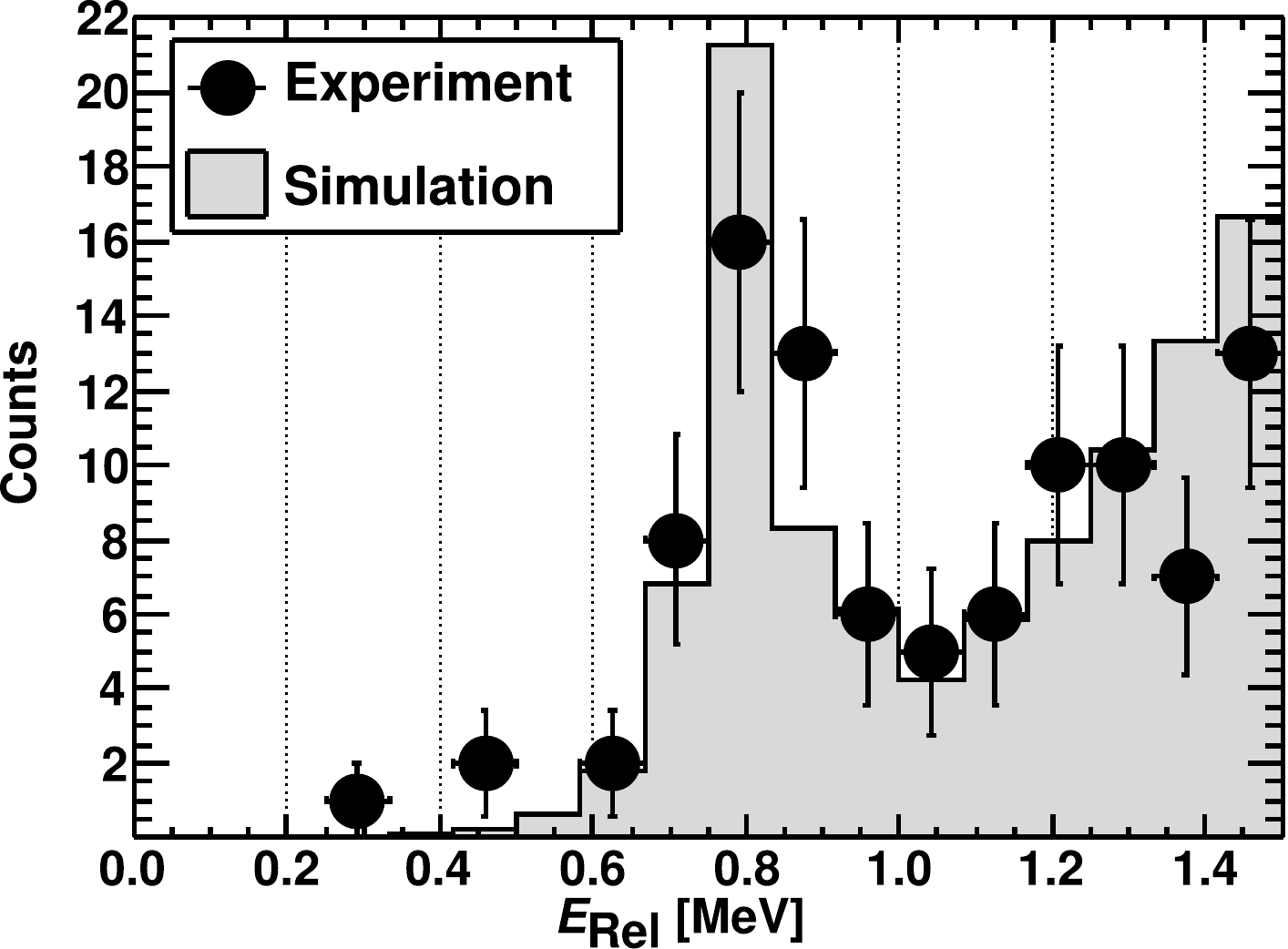}}
 \caption{(a) Relative-energy spectra for protons coincident with ${}^{68,69}$Se and ${}^{67,68}$As.  At the lowest energies, where discrete particle emission is observable, there is a distinct peak at $\sim\SI{800}{keV}$ for the reaction ${}^{69}$Br$\rightarrow p+{}^{68}$Se.  All other nuclei considered are particle-bound and therefore decay through decay modes other than particle emission.  All spectra are normalized to ${}^{69}$Br from 6-\SI{10}{MeV}. (b) Comparison of the best-fit results from a simulation using the ${}^{69}$Se mirror level ordering to the experimental data.}
 \label{fig:ERel}
\end{figure}

Figure\autoref{fig:ERelSpec} shows the reconstructed relative-energy spectrum for $p+{}^{68}$Se events (\ie\ ${}^{69}$Br proton decay) as well as comparisons to the proton emission spectra for selected neighboring particle-bound nuclei.
The ${}^{69}$Br spectrum contains two main features: a peak at $\sim\SI{800}{keV}$ and, at high ($E_{\textrm{rel}}\gtrsim \SI{1.4}{MeV}$) relative energies, a smooth distribution of proton emission events.
The protons in these events emerge through quantum mechanical tunneling from quasi-stationary states formed by the Coulomb and centrifugal barrier potentials.

The features observed in the spectra in Fig.\autoref{fig:ERelSpec} can be understood in terms of the density of states, the level widths of the decaying states, and the competition between available decay channels, namely $\gamma$ and $p$-decay modes.
For higher excitation energies, $E_x$, sufficiently above the ground state the density of states, Coulomb penetrabilities, and level widths, $\Gamma$, are increasing while the lifetimes, $\tau$, as related to the width by $\Gamma=\hbar/\tau$, are decreasing due to the reduced barrier width.
In this work states in ${}^{69}$Br are not populated selectively.
The continuous distribution observed in Fig.\autoref{fig:ERelSpec} above $\sim\SI{1.4}{MeV}$ results from proton decays from a statistical distribution of short-lived, highly energetic and overlapping resonances.
For lower excitation energies this distribution falls rapidly to zero since the density of states is lower, the proton decay lifetimes are increasing due to the increased barrier width, and other decay modes other than proton emission begin to dominate.
Particle-bound nuclei, such as ${}^{68,69}$Se and ${}^{70}$Br as considered in Fig.\autoref{fig:ERelSpec}, will decay through $\gamma$ emission while proton decay is inhibited since $\Gamma_{p}(E_{x}) < \Gamma_{\gamma}(E_{x})$.
The ${}^{69}$Br nucleus is unusual in these regards as the proton decay branch competes, and indeed dominates over the $\gamma$-decay branch since $\Gamma_{p}(E_{x}) > \Gamma_{\gamma}(E_{x})$ for the lowest states.
For the ${}^{69}$Br ground state, the competition is between $\beta$ and $p$-decay modes with $\Gamma_{p} \gg \Gamma_{\beta}$. 
At these lowest energies the density of states is decreasing and the level widths are narrow, allowing such levels to be resolved.
The peak at $\sim\SI{800}{keV}$ is consistent with these arguments and that these events originate from one-proton decay of ${}^{69}$Br.

The low-energy peak position and shape, in principle, is a combination of emission from the ground state and low-lying excited state(s) as modified by their lifetimes and by the detector resolutions.
It is assumed in the experiment that, since the lifetime of ${}^{69}$Br is short, the decay occurs at the target location.
If any of the proton-decaying states, however, are sufficiently long lived such that the decay occurs in-flight between the target and the HiRA detectors the shape of the decay peak will be modified with a tail extending toward lower relative energies.
This possibility is taken into account in the analysis using a WKB calculation.

Mirror symmetry can be used to address the structure effects on the peak shape.
There are three known low-lying levels in the $T_{z}=1/2$ ${}^{69}$Se mirror nucleus that are considered.
The mirror levels were used to generate spectra in a Monte Carlo simulation and then compared to the data.
For this analysis a spin-parity of $3/2^{-}$ is assigned to the ground state, followed by a $5/2^{-}$ level at \SI{39}{keV}, and finally a level at \SI{129}{keV} which is assigned a spin-parity of $1/2^{-}$.
In this analysis these levels are considered as pure single-particle states with unit spectroscopic factors. 
While predictions by shell-model calculations using the GX1A interaction and the systematic trend of the odd ${}^{71,73,75}$Br isotopes favor a ground-state spin-parity assignment of $5/2^{-}$ for ${}^{69}$Br this would imply a violation of mirror symmetry.
There are no known $T=1/2$ mirror nuclei where the ground state and first excited state are inverted.
Therefore, the level order of ${}^{69}$Se is used in the simulations where only the energy of the $5/2^{-}$ state is allowed to vary relative to the ground state as well as the proton separation energy.
To compare the simulation to the unbinned experimental data a Kolmogorov-Smirnov test was used.
This yields the best-fit results shown in Fig.\autoref{fig:Fit} and a proton separation energy of \mbox{$S_{p}({}^{69}\textrm{Br})={-785}^{+34}_{-40}$}.
The influence of this result on the \rp\ ${}^{68}$Se waiting-point was explored using a one-zone X-ray burst model \cite{schatz1998rpn} with reaction rates taken from the JINA reaclib database V1.0~\cite{0067-0049-189-1-240}.
At separation energies of \mbox{$S_{p}=\SI{-200}{keV}$}, \SI{-500}{keV}, and \SI{-785}{keV} the fraction of reaction flow that bypasses the waiting point by 2$p$-captures is \SI{30}{\percent}, \SI{23}{\percent}, and \SI{0.16}{\percent}, respectively.
For separation energies $\lesssim \SI{-500}{keV}$ there is a rapid reduction in flow due to the exponential dependence in the reaction rate.
With the measured separation energy of \SI{-785}{keV}, the 2$p$-capture rate has a negligible influence on mass flow through the ${}^{68}$Se waiting point in these calculations.

\section{Summary and Conclusions}
The first direct measurement of the ${}^{69}$Br proton separation energy has been made through a full kinematic measurement of the $p+{}^{68}$Se decay products.
A value of \mbox{$S_{p}({}^{69}\textrm{Br})={-785}^{+34}_{-40}$} was determined with uncertainties which allow for the 2$p$-capture rate at the ${}^{68}$Se waiting-point of the astrophysical \rp\ to be constrained.
Consequently, it is found that ${}^{69}$Br is sufficiently unbound such that ${}^{68}$Se will remain a significant waiting point in the \rp\ occurring in X-ray bursts on the surface of neutron stars.

\ack
We wish to acknowledge the support of the National Science Foundation Grant
Nos. PHY-0216783 and PHY-0855013.  

\section*{References}
\bibliography{INPC69BrDecaySimple}

\providecommand{\newblock}{}
\begin{thebibliography}{10}
\expandafter\ifx\csname url\endcsname\relax
  \def\url#1{{\tt #1}}\fi
\expandafter\ifx\csname urlprefix\endcsname\relax\def\urlprefix{URL }\fi
\providecommand{\eprint}[2][]{\url{#2}}

\bibitem{taam1993successive}
Taam R~E, Woosley S~E, Weaver T~A and Lamb D~Q 1993 {\em The Astrophysical
  Journal\/} {\bf 413} 324--332

\bibitem{schatz1998rpn}
{Schatz H \textit{et al}} 1998 {\em Physics reports\/} {\bf 294} 167--263

\bibitem{PhysRevLett.66.1571}
{Mohar M F \textit{et al}} 1991 {\em Phys. Rev. Lett.\/} {\bf 66} 1571--1574

\bibitem{Audi1995409}
Audi G and Wapstra A~H 1995 {\em Nuclear Physics A\/} {\bf 595} 409--480 ISSN
  0375-9474

\bibitem{PhysRevLett.74.4611}
{Blank B \textit{et al}} 1995 {\em Phys. Rev. Lett.\/} {\bf 74} 4611--4614

\bibitem{PhysRevC.53.1753}
{Pfaff R \textit{et al}} 1996 {\em Phys. Rev. C\/} {\bf 53} 1753--1758

\bibitem{PhysRevC.55.2407}
Ormand W~E 1997 {\em Phys. Rev. C\/} {\bf 55} 2407--2417

\bibitem{PhysRevC.65.044618}
{Lima G F \textit{et al}} 2002 {\em Phys. Rev. C\/} {\bf 65} 044618

\bibitem{PhysRevC.65.045802}
{Brown B A \textit{et al}} 2002 {\em Phys. Rev. C\/} {\bf 65} 045802

\bibitem{Wohr2004349}
{W{\"o}hr A \textit{et al}} 2004 {\em Nuclear Physics A\/} {\bf 742} 349 -- 362
  ISSN 0375-9474

\bibitem{Audi2003337}
Audi G, Wapstra A~H and Thibault C 2003 {\em Nuclear Physics A\/} {\bf 729}
  337--676

\bibitem{clark2004}
{Clark J A \textit{et al}} 2004 {\em Phys. Rev. Lett.\/} {\bf 92} 192501

\bibitem{schury:055801}
{Schury P \textit{et al}} 2007 {\em Phys. Rev. C\/} {\bf 75} 055801

\bibitem{savory:132501}
{Savory J \textit{et al}} 2009 {\em Phys. Rev. Lett.\/} {\bf 102} 132501

\bibitem{morrissey1997nhr}
{Morrissey D J and NSCL Staff} 1997 {\em NIM B\/} {\bf 126} 316--319

\bibitem{bazin2003ss}
{Bazin D \textit{et al}} 2003 {\em NIM B\/} {\bf 204} 629--633

\bibitem{yurkon1999fpd}
{Yurkon J \textit{et al}} 1999 {\em NIM A\/} {\bf 422} 291--295

\bibitem{wallace2007hra}
{Wallace M S \textit{et al}} 2007 {\em NIM A\/} {\bf 583} 302--312

\bibitem{Shapira2000396}
Shapira D, Lewis T~A, Hulett L~D and Ciao Z 2000 {\em NIM A\/} {\bf 449}
  396--407

\bibitem{Shapira2000409}
Shapira D, Lewis T~A and Hulett L~D 2000 {\em NIM A\/} {\bf 454} 409--420

\bibitem{0067-0049-189-1-240}
{Cyburt R H \textit{et al}} 2010 {\em The Astrophysical Journal Supplement
  Series\/} {\bf 189} 240

\end{thebibliography}

\end{document}